\documentclass[12pt,article]{revtex4-1}

\usepackage{graphicx,epsfig}

\newcommand{\beq}[1]{\begin{equation}\label{#1}}
\newcommand{\eeq}{\end{equation}}
\newcommand{\be}{\begin{equation}}
\newcommand{\ee}{\end{equation}}
\newcommand{\bea}{\begin{eqnarray}}
\newcommand{\eea}{\end{eqnarray}}
\newcommand{\ba}{\begin{array}}
\newcommand{\ea}{\end{array}}

\newcommand{\rarr}{\rightarrow}
\newcommand{\lsim}{\mathrel{\vcenter{\hbox{$<$}\nointerlineskip\hbox{$\sim$}}}}

\newcommand{\half}{\frac{1}{2}}
\newcommand{\nue}{\nu_e}
\newcommand{\numu}{\nu_\mu}
\newcommand{\nutau}{\nu_\tau}
\newcommand{\nuebar}{{\overline \nu}_e}

\newcommand{\Upmns}{U_{\rm PMNS}}

\newcommand{\thetaA}{\theta_{\rm A}}
\newcommand{\thetaS}{\theta_\odot}
\newcommand{\thetaR}{\theta_{\rm R}}

\newcommand{\thta}[1]{\theta_{#1}}

\begin{document}

\title{Oscillations and Mixing Among the Three Neutrino Flavors}

\author{Thomas J. Weiler}
\email[Electronic mail: ]{tom.weiler@vanderbilt.edu}
\affiliation{Department of Physics and Astronomy,
Vanderbilt University, Nashville, TN 37235, USA}

\begin{abstract}
With the educated, interested non-specialist as the target audience, 
we overview what is known and not known about contemporary neutrino physics. 
Theory tells us that neutrinos are the second-most common particle in the Universe, 
behind only the quanta of radiation called photons.  
Almost a trillion neutrinos per second enter each human eyeball,  and yet we do not see them;
these neutrinos, in roughly equal numbers, are emanations from our Sun and relics of  the hot ``big bang'' era of the early Universe.
Much of what we know about neutrinos, and hope to learn in the future, is 
derived from a unique feature of neutrinos -- ``oscillation'' among neutrino ``flavor'' types.
An initial neutrino flavor will in general oscillate into another flavor as the neutrino propagates in space and time.
Oscillations are a quantum mechanical phenomenon.   One of the wonders of neutrinos is 
that their quantum mechanics may be observed over large distances, even astronomically large.
We begin this article with neutrino phenomenology in terms of masses and mixing angles, including the matter-antimatter 
asymmetric ``phase'' that appears with three neutrino types.  
Next we venture into neutrino oscillations.  
We conclude with a discussion of model-building for the neutrino masses and mixings.  
Throughout this article, attention is devoted to the 2012 result that the 
``small'' neutrino mixing angle is not so small, after all.
\end{abstract}

\maketitle


\section{Introduction}
\label{sec:intro}
Prominent among the three thrust areas of particle physics at present are the neutrinos.
The three thrust areas are sometimes labeled the Intensity Frontier, the Energy Frontier, and the Cosmic Frontier, in the USA.
The Intensity Frontier aspires to maximize the event rate for rare processes.  
Neutrinos, characterized by some as ``as close to nothing as something can be'', are loath to interact.
Fittingly, the study of neutrinos is at the heart of the Intensity Frontier endeavor.

Incidentally, the Energy Frontier concerns itself with maximizing the energy of man-made accelerators.
This endeavor struck gold last year, with the discovery of the mysterious ``Higgs particle'' at 
the world's most energetic accelerator, the LHC located near Geneva, Switzerland.
The Cosmic Frontier studies Nature's mysteries in the cosmic setting.  
These mysteries include dark matter, dark energy, cosmic rays, and yes, more neutrinos,
this time at extreme energies.  Unlike the cosmic rays which are electrically charged and therefore bent by 
cosmic magnetic field, and unlike photons which at high energy are absorbed by ubiquitous cosmic radiation,
the charge-neutral neutrinos point back to their sources, and so have the potential to usher in the new field of 
``neutrino astronomy''.  But that is a story for another time.

Here we review what is known and not known about neutrino physics. 
Much of what we know about neutrinos, and hope to learn in the future, is 
derived from a unique feature of neutrinos -- ``oscillation'' among their ``flavor'' types.
An initial neutrino flavor will in general oscillate into another flavor as the neutrino travels.
Oscillations are a quantum mechanical phenomenon.  Quantum mechanics 
is usually observed in the domain of the very small.  However, one of the wonders of neutrinos is 
that their quantum mechanical aspect is observed over large distances, even astronomically large.
For example, the explanation of the ``anomalous'' neutrino flux observed over the Sun-to-Earth distance relies on 
quantum mechanics.
Nobel prizes in 2008, 2002, 1995, and 1988 were given in whole or part for discoveries in neutrino physics.
Conventional wisdom has it that when Nature offers a gift to scientists, it is a rare gift, and so must be savored.  
In the neutrino realm, it seems that the rarity of Nature's guiding hand has been replaced by munificence.
This overview tries to capture some of the remarkable physics that surrounds
this remarkable particle, the neutrino. 

\section{Directions and Angles}
\label{sec:angles}
At any instant of time, the position and orientation of a rotating object like the Earth, a football, or a top, 
is describable by three angles (``yaw'', ``pitch'' and ``roll'' in the language of aerodynamics) ,
measured relative to a fixed axis of $\hat{x}$, $\hat{y}$, and $\hat{z}$ directions.
These three orientation angles are called ``Euler angles''.

Why three angles, and not more or fewer?
Some thought reveals that the number of angles is specified by a rotation in each of the independent planes
of the space. Since three-dimensional space has three planes ($\hat{x}$-$\hat{y}$,
$\hat{y}$-$\hat{z}$, and $\hat{z}$-$\hat{x}$), there are three independent rotations, 
each specified by an independent rotation angle.
The final outcome of the three rotations depends on the ordering of the individual rotations.
You can prove this simply by rotating a (rectangular) book through $90^\circ$ about two axes 
in in one order, and then in reverse order, and noting that the final outcomes are different.
If we lived in two dimensions, say $\hat{x}$ and $\hat{y}$, there would be but a single plane,
the $\hat{x}$-$\hat{y}$~plane, and therefore but one rotation angle.  In four space dimensions, there
would be six rotation angles.  Continuing the count, in $N$~space dimensions there would be 
a number of planes given by the number of ways two axes may be chosen from the $N$ total axes.
Mathematicians denote this count as $C^N_2$; it is equal to $\half N(N-1)$.

Neutrino mixing works in a similar fashion.
If there are $N$ distinct neutrinos, then their distinctness defines $N$ axes.
A vector in this neutrino-space will then have components along each axis, and so describe a linear combination 
of the ``basis'' neutrinos which define the axes.  What should we take for the ``basis'' neutrinos?
Here we have to get technical.
Quantum field theory tells us that it is the distinct mass states that best describe propagation over a distance
(equivalently, the evolution of the neutrino in time - think of a movie of the motion).
So we define the basis axes to lie along neutrino mass directions,
and label the axes by the symbols $\nu_1$, $\nu_2$, etc., one axis for each neutrino.

Let us focus on the three known ``active'' neutrinos.
There my be additional neutrinos, called ``sterile'' neutrinos.
However, the evidence for additional neutrinos is not strong, 
and in some ways the evidence is contradictory,
so it is also possible that there are no more neutrinos to be discovered by our experiments.
Incidentally, the first three neutrinos are called ``active'' because they are known to participate in the 
weak interactions of particle physics.  
(Neutrinos, do not participate in the other two interactions of particle physics, the ``strong'' and 
the ``electromagnetic'' interactions.  
This is just their nature, as it is the nature of the electron to participate in the electromagnetic and weak interactions, 
but not in the strong interaction responsible for nuclear physics.)
It is further known that only three neutrinos participate in the weak interaction, so any additional neutrino species 
would interact only gravitationally, or via a new force law unique to the sterile neutrinos.
But interestingly, there is a possible mixing effect between sterile and active neutrinos, which 
could make the sterile neutrino visible to our experiments.
In this article, we shall ignore the possible existence of sterile neutrinos.
explain this possible sterile-active mixing later.

So we assume three neutrino axes, defined by the three neutrinos 
$\nu_1$, $\nu_2$, $\nu_3$, of differing mass.
These axes are the analogues of the fixed
$\hat{x}$, $\hat{y}$, and $\hat{z}$ directions of space.

What is the analog of the body-centered orientation axes?
It turns out that when neutrinos are produced by the weak interaction, such as 
occurs in decay of certain particles (charged pions, for example)
or in the fusion process inside stars and our Sun, or in supernovae explosions, or in the big-bang epoch of 
our early Universe, the produced neutrinos lie along another set of axes analogous to the body-centered frame.
Our electronics don't actually  ``see'' a neutrino. 
Commonly, at neutrino production or detection, the event transpires with an associated production or annihilation of 
a ``charged lepton'', either the electron/positron $e^\pm$, 
the muon/antimuon $\mu^\pm$, or the tau/antitau~$\tau^\pm$.
Hence, we name these interacting neutrinos at production according to their charged lepton partner, as the 
$\nue$, the $\numu$, or the $\nutau$.
It is common to call the three neutrino types, neutrino ``flavors''.
The new axes are then the direction basis in neutrino flavor space.
These neutrino flavor axes are the analog of the body-centered axes described above for the rigid-body context.
The two sets of axes, being three in number, are rotated with respect to each other by 
three ``Euler-like'' angles.

By convention, the neutrinos associated with the negative leptons are true neutrinos,
while those associated with the positive antiparticles of the negative leptons 
are the antiparticles of the neutrino, i.e.,  ``antineutrinos''.

\section{Anti-Neutrinos and a Phase}
\label{sec:phase}
For neutrinos and antineutrinos, an additional phenomenon may occur.
This phenomenon arises purely from quantum mechanics.
Quantum mechanics innately requires complex numbers for its description of Nature.
This means that, unlike the case of rotation matrices of ``classical'' physics which are 
described purely with real numbers (the Euler angles),
the rotation matrices of quantum mechanics may have complex phase factors.
Recall that complex numbers introduce the definition $i\equiv \sqrt{-1}$.
A complex phase factor is a number $e^{i\delta}$, where $\delta$ is real number
with a value in the interval $[0,2\pi ]$.
Euler's formula helps us to understand a phase factor: $e^{i\delta} = \cos\delta + i \sin\delta$.
The phase factor has a real part ($\cos\delta$) and an ``imaginary'' part ($\sin\delta$),
but always a unit length:
\beq{phaselength}
| e^{i\delta} | = |  \cos\delta + i \sin\delta | =  \sqrt{\cos^2\delta + \sin^2\delta} = 1\,.
\eeq

Let's count how many phases there are in a world with $N$~neutrinos.
We call the mixing matrix $U$.
The elements of $U$, $U_{\alpha j} \equiv \langle \nu_\alpha | \nu_j\rangle$, 
express the amount of overlap of the unit $\nu_\alpha$ neutrino flavor axis along 
the unit $\nu_j$ neutrino mass axis 
(The ``Dirac bracket''  $\langle\nu_\alpha | \nu_j \rangle$ is common notation 
for the complex-valued generalization of the dot product 
$\vec{v_1}\cdot\vec{v_2}$ which describes overlap of two real-valued vectors.)
Since $\alpha=1,...N$ and $j=1,..N$, $U$ is an $N\times N$ matrix.
An equivalent statement is that the matrix $U^*$ ``rotates'' the neutrino mass vector
$(\nu_1, ... \nu_N)$ into the neutrino flavor vector $(\nue\,\numu,\nutau,...\nu_N)$.
($U^*$ means $U$ but ``complex conjugated'', i.e., with all phases reversed in sign.)
The mixing matrix $U$ depends on 
angles and phases.
A physics constraint, that the number of neutrinos be the same when referred to either the mass axes
or the flavor axes, is that the complex-valued matrix $U$ be ``unitary'',
which has a technical definition that we do not need.
What we do need is the result that any $N\times N$ unitary matrix has $N^2$ free parameters.
Furthermore, it can be shown that with N particle pairs, $\nu_\alpha$ and charged lepton 
$\alpha,\ \alpha=1,2....N$, that $(2N-1)$ relative phases are absorbable into definitions of the 
complex-valued wave functions describing each particle.
So we are left with $N^2-(2N-1) = (N-1)^2$ physical parameters to describe neutrino mixing,
and to be determined by experiments, 
We have used $N_{\rm angles}\equiv \half N(N-1)$ of these parameters for our rotation angles in $U$.
That leaves $N_{\rm phases}\equiv (N-1)^2-\half\,N(N-1)= \half (N-1)(N-2)$ physical phases
in our mixing matrix $U$.

For a number of reasons, $N_{\rm phases}= \half (N-1)(N-2)$ is a very interesting result. 
First of all, it can be shown that the antineutrino mixing matrix is not $U$, but rather $U^*$.
For a complex-valued $U$ we have $U^*\ne U$, 
so the neutrinos and the antineutrinos mix differently.
This difference (due to nonzero phases) is being sought in our experiments.
Secondly, notice that if there were but one or two neutrino types (``flavors''),
then $N_{\rm phases}=0$, and the mixing matrix $U$ becomes  real-valued.
Thus, it is first at $N=3$ that a nonzero phase may ``complexify'' the neutrino mixing matrix.
The 2008 Nobel prize in physics was awarded for elucidation of this fact.

\section{Ordering the Rotations}
\label{sec:PDG}
Because the rotations do not commute, we must adopt a convention for 
which rotation axes are chosen and in what order.
The conventional matrix for mixing among three neutrinos, 
as established by the Particle Data Group (PDG)~\cite{PDG}, is
\bea
\label {vacPDG}  
\Upmns 
 &=& R_{32}(\theta_{32})\,U^{\dagger}_{\delta}\,
   R_{13}(\theta_{13})\,U_{\delta}\,R_{21}(\theta_{21}) \\ 
 &=&
\left(
\begin{array}{ccc}
c_{21}  c_{13} & s_{21}  c_{13} & s_{13}  e^{-i \delta}  \\ 
-s_{21}  c_{32} - c_{21}  s_{32}  s_{13}  e^{i \delta} 
& c_{21}  c_{32} - s_{21}  s_{32} s_{13}  e^{i \delta} 
& s_{32}  c_{13}  \\ 
s_{21}  s_{32} - c_{21}  c_{32}  s_{13}  e^{i \delta} & 
- c_{21}  s_{32} - s_{21}  c_{32}  s_{13}  e^{i \delta} 
& c_{32} c_{13}  \nonumber\\ 
\ea   
\right)
,.
\eea
Here, $R_{jk}(\theta_{jk})$ describes a rotation in the $jk$-plane
through angle $\theta_{jk}$, $U_{\delta}  = {\rm diag}(e^{i\delta/2},\,1,\,e^{-i\delta/2})$,
and $s_{jk} \equiv \sin \theta_{jk}$, $c_{jk} \equiv \cos \theta_{jk}$. 
The acronym PMNS stand for Pontecorvo, Maki, Nakagawa, and Sakata, all contributors 
to the early history of neutrino oscillation physics.
Note that the phase factor $e^{\pm i\delta}$ is always accompanied by the factor $\sin\theta_{13}$.
So if $\theta_{13}=0$, then the phase $\delta$ cancels from $\Upmns$ and so $\Upmns$ becomes real-valued 
(and much simpler).  We shall return to this remark later.

\begin{figure}[t]
\centering
\includegraphics[height=0.50\linewidth]{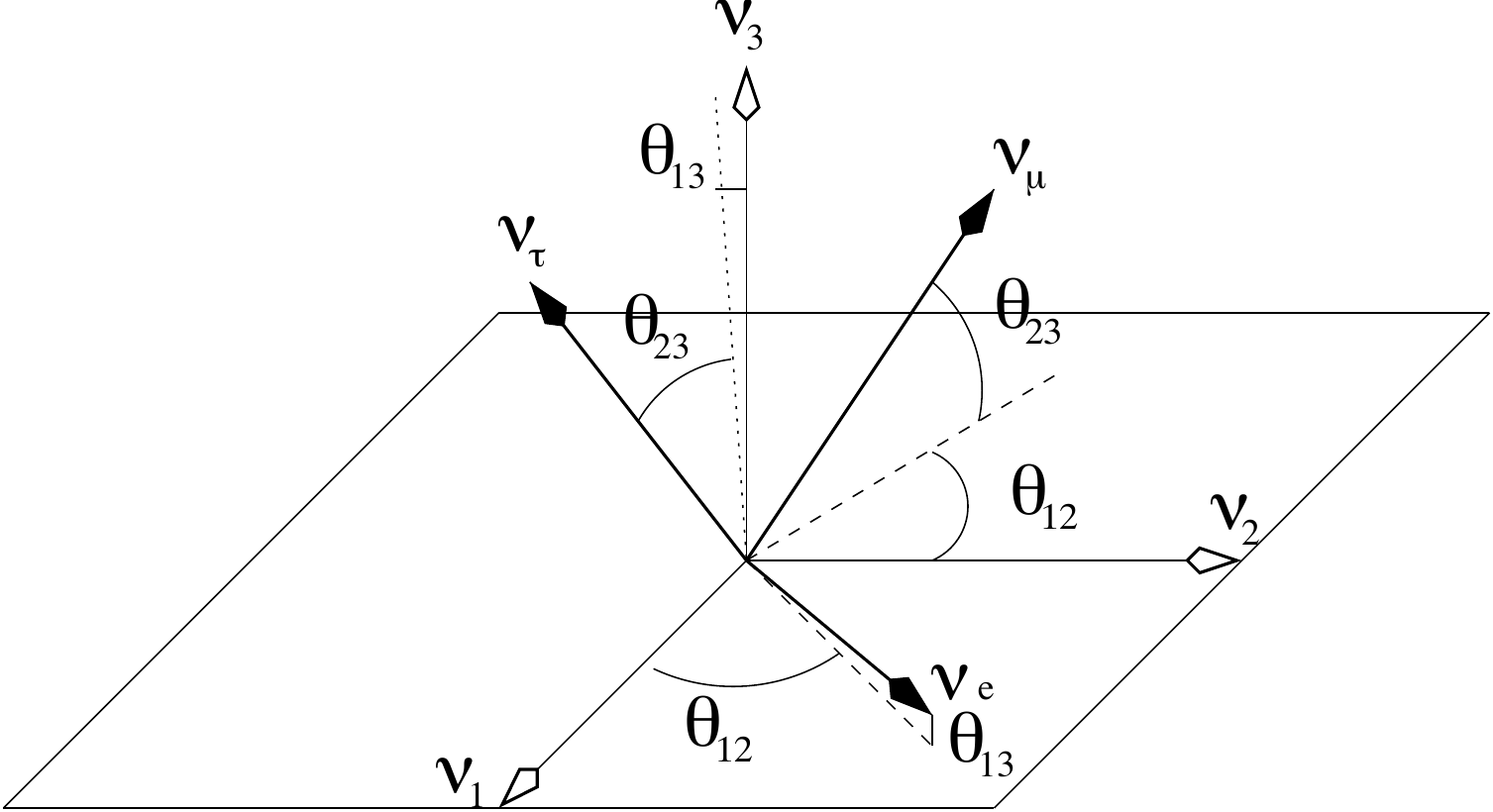}
\caption[]
{Display of the three mixing angles that characterize the orientation of the flavor axes relative to mass axes.
(figure courtesy of S. King)}
\label{fig:SKing}
\end{figure}

We have omitted two additional phases which are present only if the neutrinos are ``Majorana'' type.
These additional Majorana phases do not enter into neutrino oscillations.
Whether neutrinos are Majorana type or ``Dirac'' type present an open question and an active experimental subfield.
A Dirac fermion has four component in its wave function, to describe the two spin states 
that define a fermion such as a neutrino (we name these two components ``left-handed' and ``right-handed"), 
times the doubling implied by an independent particle and antiparticle.  
However, since the neutrino carries no strong or electromagnetic charge,
in contrast to the charged leptons and the quarks, 
there is the possibility that Nature may identify the particle with its antiparticle,
in which case just the two spin components exist in the wave function.
Interestingly, additional phases are available to the mixing matrix for Majorana neutrinos.
These additional phases arise because the neutrino-antineutrino identification fixes the phases 
of the neutrino wave functions, and in so doing disallows absorption of neutrino relative phases.
Instead of absorbing $2N-1$ phases, we absorb only $N$ phases, leaving $N-1$ additional physical phases.
For $N=3$, we get two additional Majorana phases. 
Certain processes are available to Majorana neutrino that are unavailable to Dirac neutrinos.
The process actively being sought in several experiments is ``no-neutrino double $\beta$-decay''
In double $\beta$-decay with Dirac neutrinos, one expects 
two neutrons to simultaneously decay to $2(p \nuebar e^-)$.
However, if neutrinos are their own antiparticles (Majorana particles),
the two neutrinos may effectively annihilated one another,
leading to a different, detectable final state $2(p e^-)$, occurring at a rate that depends on the 
neutrino masses, the mixing matrix $U$, and the additional two Majorana phases.
On the other hand, it can be shown that the Majorana phases do not enter into our formulae for 
neutrino oscillation, and we will not discuss them further.

\section{What We Know}
\label{sec:known}
So we have a three-active neutrino sector parametrized by 
three mixing angles ($\theta_{13}$,  $\theta_{21}$,  and $\theta_{32}$ )
and one phase ($\delta$) in $U$, and three neutrino masses ($m_1$, $m_2$, and $m_3$).
To date, all three mixing angles have been inferred from neutrino oscillation data.
In fact, because of the environments  -- nuclear reactors, the Sun, and the atmosphere -- from which their values were first deduced, these three mixing angles are sometimes referred to as $\thetaR$, $\thetaS$ and $\thetaA$.
Likewise, the mass-squared difference $\delta m^2_{21}\equiv m^2_2 -m^2_1$ 
has been inferred from oscillation data,
as has the absolute value of the difference $|\delta m^2_{32}|\equiv |m^2_3 -m^2_2|$.

The neutrino mixing angles are shown geometrically in Fig.~(\ref{fig:SKing}), 
and their values are presented in Table~(\ref{table:one}).
The generous neutrino mixing angles $\theta_{12}$, $\theta_{32}$, and $\theta_{13}$
are 3, 20, and 50 times larger than the corresponding angles from the quark sector.
This is one of many surprising features of neutrino physics, and 
Ia very beneficial feature for neutrino experimenters. 
On the other hand, the large angles present a challenge to neutrino theorists who hope to explain them.

How does it come about that we know the absolute sign of $\delta m^2_{21}$?
It is because this mass-squared difference was inferred from solar neutrino observations;
in the Sun, the background of electrons affects $\nue$ differently than $\numu$ and $\nutau$.
We do not have the space here to explain this subtle effect, but we note that it 
tells us that the lighter of $\nu_1$ and $\nu_2$ must contain more $\nue$ in vacuum (free space),
while the heavier of the two mass states must contain more $\nue$ as it emerges from the Sun.
These facts fix the mass ordering to be $m_1 < m_2$.

\begin{table}
\centering
\begin{tabular}{||c|c|c||}\hline\hline
 & {\bf Neutrinos (PMNS)} & \bf{Quarks (CKM)} \\ \hline
 $\ \theta_{12}\ $ & $35^\circ$ & $13^\circ$ \\ \hline
 $\ \theta_{32}\ $ & $43^\circ$ & $2^\circ$ \\ \hline 
 $\ \theta_{13}\ $ & $9^\circ$ & $0.2^\circ$ \\ \hline 
 $\ \delta\ $ & unknown & $68^\circ$ \\ \hline\hline
 \end{tabular}
\caption{Best-fit values of neutrino 
mixing angles compared to quark mixing angles 
(CKM denoting more historical researchers, Cabibbo, Kobayashi, and Maskawa) .
We note also that from oscillation data, 
$\delta m^2_{21}= 0.8\times 10^{-4}{\rm eV}^2$ 
and $|\delta m^2_{32}|= 2.5\times 10^{-3}{\rm eV}^2$ 
for the neutrino sector.
\label{table:one}
}
\end{table}

\begin{figure}[t]
\centering
\includegraphics[height=0.40\linewidth]{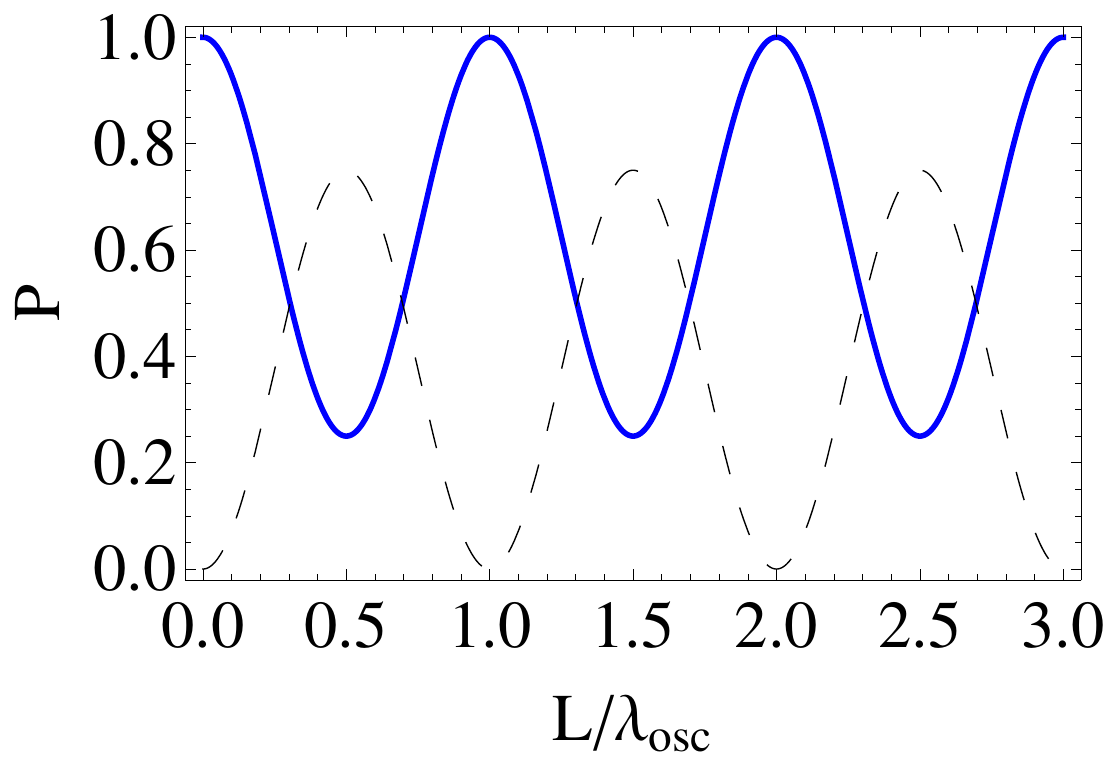}
\caption[]
{Example of two-flavor oscillation, with initial $\numu$ (blue curve) oscillating to $\nue$ (dashed black curve).
The mixing angle is $\theta=30^\circ$, i.e. $\sin^2(2\theta)=\frac{3}{4}$. (Figure courtesy of Lingjun Fu)}
\label{fig:oscn}
\end{figure}

\section{What We Don't Know -- Yet}
\label{dontknow}
Completely unknown at present are the single phase $\delta$, 
related to any neutrino-antineutrino asymmetry,
and the sign of $\delta m^2_{32}$.
The absolute neutrino masses $m_3$ and $m_2$ are of course bounded to be at least 
$\ge \sqrt{\delta m^2_{32}} \sim 0.05$~eV and $\sqrt{\delta m^2_{21}}\sim 0.01$~eV, 
but the mass ordering of these two masses is unknown.
Whether $m_3 > m_2$ (called the ``normal mass hierarchy''), 
or whether $m_3 < m_1$ (called the ``inverted'' mass hierarchy)
is a central issue in neutrino physics.
Direct searches for neutrino masses as kinks in the energy spectrum of the electron from tritium decay 
yield an upper bound $m_j \lsim$~eV, while arguments from the growth of large-scale structure in the early Universe 
yield the upper bound $\sum_j m_j \lsim 0.5$~eV.
Clearly, the three active neutrino masses are very light compared to all other known massive particles, but nonzero 
(except perhaps for the lightest neutrino).
Even the tiny electron mass is 511~keV, a million times or more than that of the neutrino,
while the mass of the arguably fundamental top-quark is almost another factor of a million larger.

\section{The Oscillation Phenomenon}
\label{sec:oscn}
Oscillations in time or distance between two states nearby in energy is a common quantum mechanical (QM) effect,
with a description available in any QM textbook.
For a two-flavor system, the probability for initial flavor state $\alpha$ to survive, 
rather than to oscillate to the second state $\beta$, 
is given by 
\beq{oscn}
P_{\alpha\rarr\alpha} = 1 - \sin^2(2\theta)\,\sin^2\left(\frac{L\,\delta m^2}{4E}\right) = 1-P_{\alpha\rarr\beta}\,.
\eeq
Here, $E$ is the neutrino energy, $\delta m^2$ is the mass-squared difference, 
and $L$ (or $t=L/c=L$ in our units) is the distance from the source.
Notice that the first $\sin^2$ with mixing angle as its argument governs the size of the oscillation,
while the argument of the second $\sin^2$ establishes the oscillation length as $\lambda = 4\pi E/\delta m^2$.
Thus, a measurement of oscillations can infer both the mixing angle $\theta$ 
and the mass-squared difference $|\delta m^2|$.
These results generalize to three-flavor systems, although the formulae become more complicated.

The origin of the oscillating $\sin^2$ terms in Eq.~(\ref{oscn}) may be explained by the following arithmetic 
(generalized to include ``imaginary'' as well as ``real'' numbers):
The Schr\"odinger equation for the propagating neutrino is $i\partial_t \Psi=E\Psi$.  
In the rest frame of a freely propagating neutrino, 
$E$ is replaced by the neutrino mass 
(Remember Einstein's $E=mc^2$? In our choice of units, c is set to unity, and so at rest, $E=m$.), 
and we label the rest frame time by $\tau$.
Then the solution to the Schr\"odinger equation is simply $\Psi(\tau) = \Psi(0) e^{-im\tau}$.
According to QM, probabilities are obtained by squaring wave-function overlaps.
So the ``survival probability'' of the flavor neutrino $\nu_\alpha$, which is a linear combination 
($\cos\theta\,\nu_1+\sin\theta\,\nu_2$) of mass neutrinos,
is the square of the ``then''-``now'' overlap:
\beq{overlapamp} 
\langle \cos\theta\,\nu_1 e^{-i \tau m_1}+\sin\theta \,\nu_2  e^{-i \tau m_2} |  \cos\theta\,\nu_1 +\sin\theta \,\nu_2 \rangle\,.
\eeq
Orthogonality of the neutrino axes is expressed as 
$\langle \nu_1 | \nu_1\rangle = \langle \nu_2 | \nu_2\rangle = 1$,
and $\langle \nu_1 | \nu_2\rangle = \langle \nu_2 | \nu_1\rangle = 0$
(just as $\hat{x}\cdot\hat{x} = 1$, $\hat{x}\cdot\hat{y} = 0$, etc.).
And so
\bea
\label{overlapprob}
P_{\alpha\rarr\alpha} &=& |  (\cos^2\theta\,e^{-i \tau m_1} + \sin^2\theta\,e^{-i \tau m_2}) |^2 \nonumber 
\\
	&=& \cos^4\theta+\sin^4\theta +2\sin^2\theta\cos^2\theta\,\Re(e^{-i\tau (m_2-m_1)})       \nonumber
\\
	&=& (\cos^2\theta+\sin^2\theta)^2 - 2\sin^2\theta\cos^2\theta (1-\Re(e^{-i\tau (m_2-m_1)}))	\,.
\eea
In the lab frame, $t=\tau/\gamma \approx \tau (m_1+m_2)/2E$ from special relativity. 
Then, setting $t=L/c=L$ in ``natural'' $c=1$ units, we get 
\beq{tautology1}
\Re(e^{-i\tau(m_2-m_1)}) = \cos(\tau(m_2-m_1)) \approx \cos(L\,\delta m^2/2E)\,,
\eeq
 and so with the use of a trigonometric identity, we arrive at 
\beq{tautology2}
1-\Re(e^{-i\tau (m_2-m_1)}) = 2\sin^2(L\,\delta m^2/4E)\,.
\eeq
Setting $ (\cos^2\theta+\sin^2\theta)^2 $ equal to one, and using Eq.~(\ref{tautology2}), 
we arrive at Eq.~(\ref{oscn}).
  
\section{The Newest Discovery -- Large {\boldmath$\theta_{13}$} -- and Its Implications}
\label{sec:theta13}
For many years, the sparse data on the angle $\theta_{13}$ allowed consistency with zero.
However, in spring of 2012, $\theta_{13}$ was definitively measured to be nonzero, 
and by a large margin~\cite{DayaBay,RENO}.
This discovery of a ``large'' $\theta_{13}$ has several consequences.
Four that we will discuss briefly here are {\bf (i)} the increased reach for experiments to infer the 
neutrino mass hierarchy;
{\bf (ii)} the significantly larger amount of particle-antiparticle asymmetry (called ``CP violation'')
in the theory (which could explain the the asymmetric fact that we now live in a matter-dominated Universe,
rather than a matter-antimatter symmetric Universe); {\bf (iii)} the breaking of $\numu$-$\nutau$ symmetry;
and {\bf (iv)} the increased difficulty of models to accommodate the large observed mixing angles.

Point {\bf (i)} is explained in that oscillations in vacuum and in matter are enhanced by larger mixing angles.
Point {\bf (ii)} is more subtle.  It turns out that the parametrization-independent measure of CP~violation in 
the three-neutrino system is given by 
$J\equiv | \Im(U_{\alpha j}U^*_{\beta j}U_{\beta k}U^*_{\alpha k}) |$, for any $\alpha\ne\beta$, and any $j\ne k$.
With three neutrinos, $J$ has the same value for any choice of these indices,
a fact related to the one and only phase in the mixing matrix.
In terms of the PDG parameters given in Eq.~(\ref{vacPDG}), one finds
\beq{J}
 J=\frac{1}{8}\,\cos\thta{13}\,\sin(2\thta{13})\,\sin(2\thta{12})\,\sin(2\thta{32})\sin\delta\sim 0.036\sin\delta\,.
 \eeq
 Notice that all three angles and the phase must be nonzero for $J$ to be nonzero.
 We already saw that a zero value for $\theta_{13}$ implies a real-valued $\Upmns$,
 which in turn implies a vanishing value for $J$.  Eq.~(\ref{J}) generalizes that finding. 
 Notice that $J$ grows as $\sin(2\theta_{13})$, so as $\theta_{13}$ moves away from zero, $J$ increases.
  One may infer the robustness of this result by contrasting $J$ with the analogous quantity in the quark sector, 
 $J_Q\approx 0.30\times 10^{-4}$.  We have $J\approx (1000\,J_Q) \, \sin\delta$.
 Of course, $\sin\delta$, completely unknown at present, must be nonzero for $J$ to be nonzero.
 
Point {\bf (iii)} concerns the breaking  of $\numu$-$\nutau$ symmetry.
When $\theta_{32}$ is set to the maximal mixing value of $45^\circ$ and $\theta_{13}$ is set to zero,
a significant increase in symmetry of the mixing matrix arises.
The norms $|U_{\mu j}|$ become equal to the norms $|U_{\tau j}|$, for each $j=1,2,3$.
This is the mathematical statement of what is termed $\numu$-$\nutau$~symmetry.
If, in addition, the norms of all second column elements $|U_{\alpha 2}|$, $\alpha=e,\mu,\tau$,
are set to be identical, the symmetry is increased further.
This further symmetry requires that $\theta_{21}=\tan^{-1}(\frac{1}{\sqrt{2}})$.
This version of $\numu$-$\nutau$~symmetry came to be known as ``tri-bimaximal mixing'' (TBM),
the unwieldy name reflecting some vestigial history.
The TBM mixing matrix is 
\beq{TBM}
U_{\rm TBM}=
R_{32}\left(\theta_{32}=\frac{\pi}{4}\right)\,R_{13}(\theta_{13}=0)\,R\left(\theta_{21}=\tan^{-1}\left(\frac{1}{\sqrt{2}}\right)\right)
= \frac{1}{\sqrt{6}}
\left(
\ba{rrr}
2 & \sqrt{2} & 0 	     \\
-1 & \sqrt{2} & \sqrt{3} \\
1 & -\sqrt{2} & \sqrt{3} \\
\ea
\right) \,.
\eeq
When $\theta_{13}$ was thought to be zero or nearly so, 
$U_{\rm TBM}$ was the natural and popular choice for the zero$^{\rm th}$~order
matrix about which perturbations could be added to better accommodate data.

Although the values of the mixing angles depend on the ordering of the Euler-like rotations in neutrino space,
the norms of the matrix elements of $U_{\rm PMNS}$ (here, $U_{\rm TBM}$) are convention-independent.
This means that the magnitudes of the matrix elements may reveal some underlying physics which is not presently known.
It is intriguing that the values of the matrix elements are among those that arise from quantum-mechanical addition of 
two angular momenta, spin and/or orbital.  
(The quantum-mechanical addition factors are called ``Clebsch-Gordon coefficients'' after 
two $20^{th}$~century physicists.)  
It is intriguing to speculate that these familiar  
addition coefficients may indicate another inner layer of the particle onion, namely a composite nature of neutrinos -- 
that two more-fundamental particles might combine to form a bound state which we call the neutrino, 
with the individual particle properties adding to give the measured properties of the neutrino.
At present, I am unaware of any model that support this conjecture;
it is difficult to contemplate bound states with masses as light as the neutrino 
when there is no experimental evidence of any composite substructure all the way up to 
the energy scale of the weak interaction.

Incidentally, the equality of elements in the middle column of $U_{\rm TBM}$ 
implies an equality of flavors in the mass state $\nu_2$.
This equipartition of neutrino flavor among the $\nu_2$ state resolved the ``solar anomaly'',
which had vexed neutrino and solar physicists for over twenty years!
It turns out that because of the smoothly varying electron density from the Sun's core to its corona, 
the solar neutrinos which are produced in the fusion cycle of the core as pure flavor states $\nuebar$, 
emerge from the Sun as almost pure mass states $\nu_2$ (via subtle QM called the ``MSW'' effect).
Thus, the solar neutrinos are perceived on Earth as very nearly equal mixtures of 
$\nuebar$, $\bar{\nu}_\mu$, and $\bar{\nu}_\tau$,
i.e., the measured $\nuebar$ flux at Earth is only 1/3  of the original solar $\nuebar$ flux.

Now that $\theta_{13}$ is known to be $\sim 9^\circ$, and 10 standard deviations removed from zero,
the $\numu$-$\nutau$ symmetry of TBM is broken (barring special values of the unknown phase $\delta$).
This leaves model-building of the neutrino masses and mixing in a much more complicated and confused state,
encapsulated in our point~{\bf (iv)} above.  
In particle physics we are accustomed to breaking a symmetrical situation 
``perturbatively'',  meaning with small asymmetries.
The inferred value of $\theta_{13}\sim 9^\circ$ is not perturbatively small.
The $\numu$-$\nutau$ symmetry of TBM is rather badly violated.
One possibility is to introduce large perturbations.
The jury is out on whether such a direction will bear any fruit,
and one is right to ask if the TBM basis with subsequent large symmetry breaking should be replaced by 
some other, more symmetric basis that accommodates smaller symmetry breaking 
than does the TBM model.
Again, the jury is out on the fruitfulness of the ``new basis'' approach.

\section{Neutrino Model-Building -- and Some Group Theory}
\label{sec:modeling}
Generally, the neutrinos' kinetic energy terms and interaction terms are more symmetric than the mass terms;
the latter therefore provide symmetry-breaking information.  
Diagonalization of the mass matrix produces the transformation from the interaction or flavor axes to the mass axes,
i.e., provides the mixing matrix $\Upmns$.
The resulting mixing matrix generally shows some residual symmetry, but also some significant symmetry breaking.
For example, the TBM mixing pattern of Eq.~(\ref{TBM}) follows naturally from diagonalization of the four-parameter 
mass matrix
\beq{TBMmass}
M_{\rm TBM}=
\left(
\ba{ccc}
\mu_1 & \mu_2 & \mu_2    \\
\mu_2 & \mu_3 & \mu_4    \\
\mu_2 & \mu_4 & \mu_3    \\
\ea
\right) \,.
\eeq
This mass matrix has $\numu$-$\nutau$~symmetry, as shown by invariance under simultaneous interchange  
of the 2nd and 3rd rows and 2nd and 3rd columns (the $\numu$ and $\nutau$ rows and columns).
This $\numu$-$\nutau$~symmetry is reflected in the $U_{\rm TBM}$ mixing matrix of Eq.~(\ref{TBM}).

A discussion of flavor symmetry and symmetry breaking necessarily invokes the mathematics of group theory
and group representations.
Suppose the neutrinos were massless, or mass-degenerate, meaning that all neutrinos have a common nonzero mass.
Then interchange of the neutrinos, or rotation among them, could not change the physics.
The collection of all three-dimensional spatial rotation operations is the group is named $SO(3)$.
The meaning of the ``3'' is obvious, the ``O'' stands for ``orthogonal'', and the ``S'' signifies unit determinant. 
However, quantum mechanics is intrinsically complex-valued rather than real-valued, 
and so the rotations in neutrino space are unitary rather than orthogonal,
leading to the ``special unitary group'' $SU(3)$,
characterized by $3\times 3$ unitary matrices having unit determinant.
The unit determinant constraint removes one parameter, leaving an $N^2-1\stackrel{N=3}{\rarr}8$-parameter group
(which may be decomposed into the eight Gell-Mann matrices, but that is another story for another time.)
The eight-parameter group $SU(3)$ is then the symmetry group of the three flavors, 
before any symmetry breaking due to mass differences is admitted.

In the Standard Model (SM) of particle physics, the active neutrinos and their charged lepton partners
must rotate together under each $SU(3)$ rotation.
When mass differences and off-diagonal terms in the mass matrix of the 
neutrinos and/or charged leptons are admitted, 
the symmetry of the $SU(3)$ group is said to be ``broken''.
Diagonalization of the charged lepton mass matrix is accomplished with a unitary matrix called $V_{\ell^\pm}$.
Diagonalization of the neutral lepton (``neutrino'') mass matrix is accomplished with a unitary matrix called $V_\nu$
(If the neutrinos are of the Majorana type, then the mass matrix can be shown to symmetric,
and diagonalization proceeds via $V_\nu M V_\nu^{\rm T}$ rather than via $V_\nu MV_\nu^\dag$.)
If the two matrices $V_{\ell^\pm}$ and $V_\nu$ were identical, 
then the common unitary rotation would just reflect the $SU(3)$ invariance of the mathematics (what physicists call the ``Lagrangian''),
and not to any physical effect.
However, if the two matrices are different, then there is physics involved.
The invariant mixing matrix (up to some quantum mechanical phases) is the misaligned product 
%
$\Upmns \equiv V_{\ell^\pm} V_\nu^\dag$.
%
Since overall rotations are not physical, it is convenient, allowable, and common, 
to work in a rotated basis where the charged lepton matrix is already diagonal.
But the underlying physics is that symmetries of the massless Lagrangian may be broken in the 
charged lepton sector via $V_{\ell^\pm}$ or $V_\nu$, or more likely, both.

In the language of group theory, we say that the initial symmetry, here $SU(3)$, is broken by masses to a
subgroup $G_{\ell^\pm}$ for the charged leptons, and a subgroup $G_\nu$ for the neutrinos.
Then $\Upmns$ results from the mismatch of the way Nature breaks the large $SU(3)$ symmetry 
to the two subgroups $G_{\ell^\pm}$ and $G_\nu$.
The full continuous group $SU(3)$ offers many breaking patterns that fail 
to constrain the resulting mixing angles.
Recent models constructed to ``explain'' the two large and one small mixing angles of the neutrino sector 
use discrete subgroups of $SU(3)$ for $G_{\ell^\pm}$ and $G_\nu$.
The smaller discrete subgroups such as $S_3$, $S_4$, $A_4$, $A_5$, result in an over-constrained system of 
mixing parameters, and therefore predict mixing angles or relations among mixing angles.
Individual particles are assigned to a ``group representation'', 
a kind of flavor vector whose members rotate among themselves under general group rotations.
Each distinct group has its own unique set of ``representations''.
The small discrete groups contain several singlet representations for particles, 
and some doublet and triplet representations,
perfect for assignments of the three active neutrinos and three charged leptons.
Once a particular discrete flavor group is chosen and particles are assigned to the group's representations, 
the flavor group is broken by assigning large vacuum expectation-values (``vevs'') to flavor-scalars 
(called ``flavons'' or ``familons''), into a neutrino sector group $G_\nu$ and a into charged-lepton sector group $G_{\ell^\pm}$. 
The small groups including those just mentioned tend to contain an inherent $\numu$-$\nutau$ symmetry, 
which as we have seen leads to the prediction that $\theta_{13}$ is zero,
or in the case of natural choices for symmetry breaking, nearly zero.
Consequently, the recent measurement of a large-ish value for $\theta_{13}\sim 9^\circ$,~\cite{DayaBay,RENO}
presents a challenge to the implementation of these small discrete subgroups.

An alternative approach is to embrace the larger discrete subgroups of $SU(3)$.
Larger groups present larger representations for the particle assignments,
and tend at leading order (i.e., without perturbations) to get $\theta_{13}$ about right.
$\Delta(96)$ and $\Delta(384)$ are examples in current use.
(Here the argument of the discrete group refers to the number of elements in the group.)
While these groups may appear ``big'' compared to the first generation of ans\"atze,
we should not summarily dismiss particular groups as being ``too large'' --  Nature, 
the ultimate arbiter, may have chosen one of them to fit ``just right''.

A consistency check on the validity of some discrete groups is provided by the model's predictions for rare 
flavor changing processes, such as $\mu\rarr e\gamma$ and $\mu\rarr 3e$.
In the SM, such branching ratios of the muon are proportional to $m_\nu^2$, and so negligibly small.
But Beyond the SM, such branching ratios may be observable, due to enhancements from the flavor-scalars.
The MEG experiment~\cite{MEG} has attained an upper limit of the branching ratio $\mu\rarr e\gamma$ of $5\times 10^{-13}$.
A reach to $5\times 10^{-14}$ is expected within this decade.
Also, at Fermilab in the US~\cite{Mu2e}, and at JPARC in Japan~\cite{Comet}, dedicated experiments to search for 
$\mu\rarr e$ transitions in nuclei are presently under construction.
Results from these experiments, now and in the future, (will) invalidate or constrain particular neutrino-mixing models. 

\section{Summary}
\label{sec:summary}
In this pedagogical article, we have overviewed some experimental and theoretical aspects of one of the most exciting 
arenas in present-day particle physics, that of the neutrino.
The three angles and one phase that characterize the misalignment of the neutrino ``flavors'' and neutrino masses were introduced.
These parameters, and the neutrino masses themselves, were shown to enter and emerge 
from neutrino oscillation studies.  The phenomenon of neutrino oscillations is quite interesting in its own right,
being a macroscopic manifestation of quantum mechanics at work.

The recent inference of $\theta_{13}$ completes are knowledge of the three neutrino mixing angles.
We discussed how the fortuitously large value of $\theta_{13}$ increases the reach of experiments to reveal 
the ordering of the neutrino masses (the ``mass hierarchy''), 
and to discover matter-antimatter asymmetries ($CP$ or $T$~reversal violation) in the neutrino sector.
On the other hand, as we discussed, large  $\theta_{13}$ also complicates the building of group-theoretic models
that are presumed to underlie the neutrino mixing angles and mass values.

Neutrino physics is very much an ongoing enterprise, both experimentally and theoretically.
If the past is any guidepost to the future,
we can expect surprising results from the continued search into the nature of neutrinos, their masses, mixings, and interactions.



\begin{thebibliography} {99}

\bibitem{PDG} The Particle Data Group, accessible online at 
pdg.lbl.gov





  
\bibitem{DayaBay}  F.~P.~An {\it et al.}  [Daya Bay Collaboration],
  Chin.\  Phys.\ C {\bf 37}, 011001 (2013)
  [arXiv:1210.6327 [hep-ex]];
  ibid,  
  Phys.\ Rev.\ Lett.\  {\bf 108}, 171803 (2012)
  [arXiv:1203.1669 [hep-ex]].

\bibitem{RENO}  J.~K.~Ahn {\it et al.}  [RENO Collaboration],
  Phys.\ Rev.\ Lett.\  {\bf 108}, 191802 (2012)
  [arXiv:1204.0626 [hep-ex]].


\bibitem{MEG} MEG webpage at meg.web.psi.ch

\bibitem{Mu2e} Mu2e webpage at mu2e.fnal.gov

\bibitem{Comet} 
  Y.~Kuno [COMET Collaboration],
  PTEP {\bf 2013}, 022C01 (2013).


\end{thebibliography}
\end{document}